\documentclass{article}
\usepackage{graphicx}
\usepackage{sidecap}
\usepackage{wrapfig}
\usepackage{setspace}
\usepackage{rotating}
\usepackage{amsmath}
\usepackage{endnotes}
\usepackage[left=3cm,top=3cm,right=3cm,bottom=3cm]{geometry}
\usepackage{overcite} 
\usepackage{appendix}
\usepackage{chngcntr}
\usepackage{soul}
\usepackage{pdflscape}
\usepackage{lineno}
\usepackage{caption}
\usepackage{enumerate}
\usepackage{authblk}

\newcommand{\beq}{\begin{equation}}
\newcommand{\eeq}{\end{equation}}

\newcommand{\bi}{\begin{itemize}}
\newcommand{\ei}{\end{itemize}}

\title{Attributing excess conflict risk in Syria to anthropogenic climate change}

\author[]{Solomon Hsiang and Marshall Burke\thanks{Stanford University and the National Bureau of Economic Research; solhsiang@stanford.edu, mburke@stanford.edu}}

\date{}
\begin{document}

\maketitle
\singlespacing

\begin{abstract}
\noindent Some analyses suggest that drought in Syria during 2007-10 may have contributed to the 2011 onset of its civil war and that anthropogenic climate change had a detectable impact on the severity of that drought. Yet these qualitative statements alone do not allow us to estimate {how much} anthropogenic climate change amplified the risk of civil war in Syria. Climate policy is increasingly relying on global-scale cost-benefit analyses to inform major decisions, and any excess conflict risk imposed on future populations should be considered in this accounting because of its substantial human and economic impact.  In order for this excess conflict risk to be accounted for, it must be quantified and its uncertainty characterized.  Here we build on multiple recent findings in the literature to construct a best estimate for the excess conflict risk borne by Syria in 2010 that was attributable to anthropogenic climate change. We estimate that the baseline risk of conflict in Syria, which was likely already high, was amplified roughly 3.6\% (90\% confidence interval: 1.1-7.3\%) due to the anthropogenic component of the drought.  The effect of climate change was thus discernible but unlikely responsible for the bulk of Syrian conflict risk.  Nevertheless, the magnitude of similar excess risk around the globe is expected to grow as climate change progresses, and its human cost could be large because most populations will face positive excess risk.  Our approach to quantifying this risk is applicable to these other settings. 

\end{abstract} 

\newpage


Archeological and paleoclimatic evidence suggest that the implosion of the Akkadian Empire in the fertile crescent (modern day Syria and Iraq) may have been triggered by regional drying four millennia ago \cite{Cullen2000}. Work by Kelley et al (PNAS, 2015) demonstrates that anthropogenic climate change is likely responsible for intensifying the modern drought in the same region \cite{kelly2015}. The authors argue that because the drought caused agricultural failures, rapid migration to urban areas, strains on public resources, and social unrest,  it is likely that anthropogenic climate change contributed to the initiation of ensuing full-scale civil war. History may be repeating itself, albeit this time with a human fingerprint on the climate change dial.

To date, researchers have used quasi-experimental approaches to quantify how abnormal climatic events, usually abrupt ones, have historically increased the risk of conflict and violence in human societies around the globe \cite{Science2013}.  However, no study has attributed a quantitative change in conflict risk to observable anthropogenic climate changes. Kelley et al. estimate how much stronger the 2007-10 Syrian drought was relative to a counterfactual with no anthropogenic climate change, but they stop short of computing the additional conflict risk caused by this strengthening. Here we build on the results of Kelley et al. and earlier findings \cite{Science2013, ARE2015} to construct an estimate for the \emph{ex ante} excess conflict risk attributable to the anthropogenic component of the drought. Essentially, this is the additional risk an observer would have assigned to Syria if they were standing at an earlier moment in time, say the year 2000, with the information that human activity would exacerbate the future drought (as published in Kelley et al) and that drought increases conflict risk (as in \cite{Science2013, ARE2015}).  This calculation is analogous to how one might compute excess conflict risk today for future populations subject to additional future climate change. 

Every conflict is unique, making precise forecasts of conflict extremely difficult. Nonetheless we have identified systematic patterns across conflicts \cite{Science2013} that can help us estimate \emph{ex ante} risk even if we do not know all the contextual details of a population and cannot resolve all future events relevant to a potential conflict.  Such attribution of risk does not require exactly predicting whether an event will occur, rather it is an assessment of the change in probability that an event might occur. Society regularly undertakes similar risk assessments and uses them to improve wellbeing. For example, laws discourage driving while consuming alcohol and flights are regularly grounded in extreme weather. In both cases, specific factors elevate the risk of a deadly accident enough that we encourage individuals to change their behavior, even though the contextual details and specific events related to a specific potential crash are unknowable \emph{ex ante}. 

To attribute excess conflict risk to observable anthropogenic climate changes, we combine the results from Kelley et al. with global average estimates describing the effect of drought on conflict risk \cite{ARE2015}.  Data describing numerous perfect historical duplicates of Syria under drought and no-drought conditions do not exist, preventing the construction of a Syria-specific estimate for this effect (similar to the difficulty in constructing driver-specific accident risk due to alcohol consumption).  Instead we use the best known numbers applied to this context, thereby assuming Syria behaves similar to a randomly selected population from elsewhere in the world \cite{ARE2015}.  This represents the best estimate, given available data.  Should future evidence indicate that Syrians are fundamentally different than all other populations, then these estimates should be adjusted. Nonetheless, these calculations represent an important exercise because they enable us to begin including conflict impacts in formal risk assessments and cost-benefit analyses of climate change.  Furthermore, the discipline of formal risk calculation prevents us from over- or under-exaggerating the role of climate change in Syria and other contexts.

Our approach is mathematically similar to attribution in the agronomic literature \cite{lobell2011}, and has immediate applicability in other settings where the overall relationship between a climate variable and an outcome of interest has been described and where the anthropogenic component of an observed climate event can be detected. 

Even if no anthropogenic climate change had occurred, it is likely that Syria would have experienced a natural drought in 2007-10 \cite{kelly2015} and this drought would have increased the risk of intergroup conflict relative to a scenario with no natural drought \cite{Science2013}. Thus, it is inaccurate to attribute all of the excess conflict risk caused by the drought to anthropogenic climate change. The component of excess conflict risk attributable to anthropogenic climate change is the risk of conflict due to the observed drought minus the risk that would have been borne had the drought occurred but been unaffected by climate change.

Conflict is complex and multidimensional in nature, thus we focus only on the partial effect of drought on violence, holding other aspects of the Syrian context fixed.  Let  $P$ be the probability of conflict, with a baseline probability of conflict $P_{0}$ attributable to all contextual factors independent of the 2007-10 drought, such as political institutions, history, and culture. Let the drought have a component of its magnitude $D_0$ that would have occurred naturally without anthropogenic climate change and a component $D'$ attributable to humans. The observed drought then had magnitude $D_0+D'$. Assuming the expected probability of conflict can be described by a continuous response surface $P(D)$ that is differentiable with respect to drought, then the total probability of conflict is
\begin{equation*}
P =P_{0} + \underbrace{\int_{0}^{D_{0}}\frac{\partial P(D)}{\partial D}dD}_{\mathrm{natural}} + \underbrace{\int_{D_{0}}^{D_{0}+D'}\frac{\partial P(D)}{\partial D}d D}_{\mathrm{anthropogenic}}
\end{equation*}
where the last term is the component of excess conflict risk attributable to anthropogenic climate change. Dividing through by $P_0$ we have
\begin{equation}
\frac{P}{P_0}\approx 1 + \beta D_{0} + \beta D'
\label{Eq:components}
\end{equation}
where the approximation holds if $\frac{\partial P(D)}{\partial D}\cdot\frac{1}{P_0}$ is roughly constant and approximated by $\beta$, as suggested by analysis in refs [\cite{Science2013, ARE2015}].  Note that \emph{relative risk} is linear in $D$, but drought has an increasingly large effect on the \emph{probability} of conflict as baseline conflict rates $P_0$ increase, i.e. $P(D)$ is nonlinear. 

\begin{sidewaysfigure}
\begin{center}
\includegraphics[width=\textwidth]{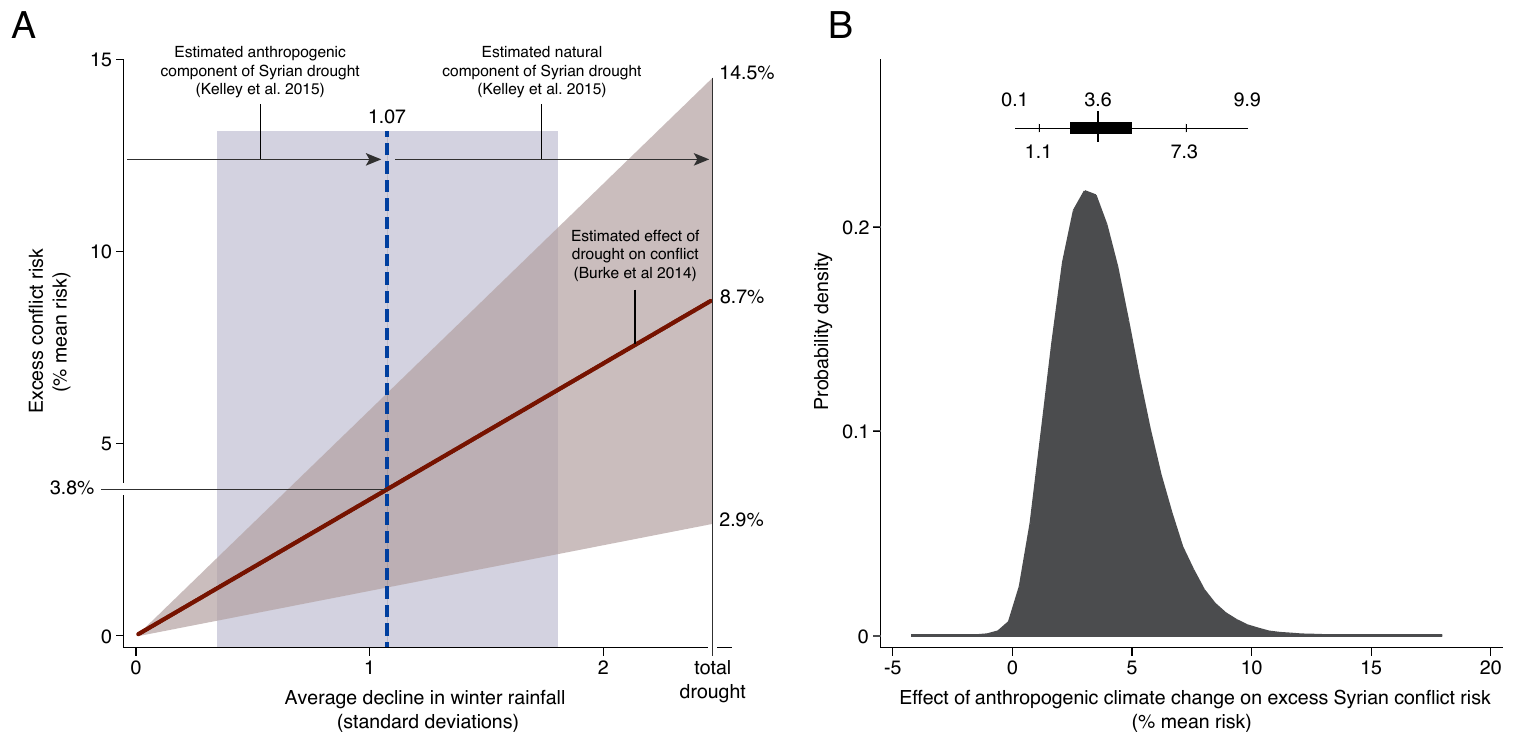}
\caption{\textbf{Attributing excess conflict risk in Syria to the component of drought caused by humans.} (A) The 2007-2010 drought (measured in standard deviations of historical variability) has a natural component induced by climate variability ($D_0$) and a trend component that has been attributed to anthropogenic climate change in ref. [\cite{kelly2015}] ($D'$, dashed blue w. 95\% CI). Ref. [\cite{ARE2015}] uses meta-analysis of 17 harmonized studies to estimate that each standard deviation in persistent drought generates 3.54\% ($\pm$1.2\%) excess conflict risk (red w. 95\% CI, slope $=\beta$). Combining these estimates via equation \ref{Eq:components} allows us to attribute a component of excess conflict risk in 2008 due to anthropogenic climate change (mean estimate: 3.8\%). (B) Combining uncertainty in the estimated size of the anthropogenic component of drought \cite{kelly2015} and uncertainty in the sensitivity of conflict to drought \cite{ARE2015}, we construct the probability distribution of excess conflict risk in 2010 Syria attributable to anthropogenic climate change. Median, interquartile range, 90-centile range and 99-centile range indicated by box-whisker.}
\label{mainfig}
\end{center}
\end{sidewaysfigure}

Kelly et al show that wintertime rainfall in 2007-10 was on average 2.48$\sigma$ below the historical mean and estimate that 44\% of this deficit ($D'=1.08\pm.37\sigma$) was due to anthropogenic climate change (personal communication). Ref [\cite{ARE2015}] updates the analysis of ref [\cite{Science2013}] using a harmonized meta-analysis and estimates $\beta = 3.54\pm1.2\%\sigma^{-1}$ for a precipitation-related climatic disturbance that persists for two consecutive observational periods. Thus the natural component of the drought likely elevated conflict risk $\beta D_0=4.9\%$ above baseline and the anthropogenic strengthening of the drought increased conflict risk by an additional $\beta D' = 3.8\%$ on average.  Figure \ref{mainfig}A displays how these two components together amplified the total risk of conflict in Syria. The figure also displays how uncertainty in the sizes of $D'$ and $\beta$ interact to generate uncertainty in their product. Accounting for both of these uncertainties, we attribute a 1.1-7.3\% increase in Syrian conflict risk due to anthropogenic climate change (90\% CI). The full probability distribution of risk estimates are shown in Figure \ref{mainfig}B.

Kelly et al document that Syria also experienced exceptional anomalous warming during 2007-2010.  This warming likely had a large effect on Syrian conflict risk in 2010, as the effect of persistent warming on conflict risk estimated by ref [\cite{ARE2015}] is substantially larger than the effect of low precipitation ($11.33\pm2.96\%\sigma^{-1}_{T}$).  However we do not account for the effect of this warming here because Kelley et al do not attempt to attribute that warming to anthropogenic climate change.  

Based on these estimates, it seems unwarranted to infer from the analysis of ref [\cite{kelly2015}] that the entirety of the Syrian conflict was due to anthropogenic climate change. Nonetheless, that anthropogenic climate change can be attributed with generating \emph{any} conflict risk is novel and we reject the hypothesis that anthropogenic climate change had no effect ($p<0.01$). The estimated 1.1-7.3\% amplification of 2010 Syrian conflict risk via anthropogenic climate change is quantitatively modest, but we caution that similar effects around the world are expected to grow with time as we continue to alter the climate.


\paragraph{Acknowledgements}
We thank Colin Kelly for data and Mark Cane, Richard Seager, Yochanan Kushnir, Michael Oppenheimer, and Tamma Carleton for comments and suggestions.\\

\newpage

\end{document}